\documentclass[aps,prl,twocolumn, groupedaddress, showpacs]{revtex4}
\usepackage{graphicx}
\usepackage{dcolumn}
\begin{document}

\title{Observation of classical rotational inertia and nonclassical supersolid signals in solid $^4$He below 250 mK}

\author{Ann Sophie C. Rittner and John D. Reppy}
\email{jdr13@cornell.edu}

\affiliation{Laboratory of Atomic and Solid State Physics and the
Cornell Center for Materials Research, Cornell University, Ithaca,
New York 14853-2501}

\date{\today}

\begin{abstract}
We have confirmed the existence, as first reported by Kim and Chan,
of a supersolid state in solid $^{4}$He at temperatures below 250
mK. We have employed a torsional oscillator cell with a square cross
section to insure a locking of the solid to the oscillating cell. We
find that NCRI signal is not a universal property of solid $^{4}$He,
but can be eliminated through an annealing of the solid helium
sample. This result has important implications for our understanding
of the supersolid state.
\end{abstract}


\maketitle Eunseong Kim and Moses Chan (KC) of Penn State University
first reported \cite{1} the observation of supersolid behavior for
solid $^{4}$He confined within the porous media of porous gold and
Vycor glass. KC followed this remarkable discovery with the
observation of a supersolid or non-classical rotational inertia
(NCRI) signal in bulk solid helium \cite{2}. An important goal,
which we have achieved in our experiments, has been to provide an
independent confirmation of the KC observations for bulk $^{4}$He.
We have also investigated the influence of annealing and sample
preparation on the existence of the supersolid state. We find that
the supersolid state is not a universal property of solid $^{4}$He,
but that it is possible through crystal annealing near the melting
curve to create samples that show only classical rotational inertia
(CRI).

In these measurements we have employed the torsional oscillator
technique developed at Cornell over the past several decades
\cite{3}. Oscillators with two different sample geometries were used
in the experiments reported in this letter. The first had a
cylindrical sample geometry with an internal volume of 2.7 cm$^{3}$
and operated at a frequency of 253 Hz, while the second had a cubic
geometry with a volume of 1.4 cm$^{3}$ and operated at 185 Hz. The
helium used in the experiments was commercial well-grade helium
similar to that used by KC in their bulk solid helium work \cite{2}.
This helium has a stated $^3$He impurity level of 0.2 - 0.3 ppm.

Our earliest runs were made with the cylindrical cell. Data obtained
with this cell at a pressure of 27 bar showed an onset of the
supersolid signal at a temperature near 0.2 K, with a maximum signal
amplitude at 30 mK of 0.6 \% of the total period shift seen upon
forming the solid in the cell. The magnitude of this signal is in
good agreement with the observations of KC made under similar
conditions and provides a confirmation of their results. In an
attempt to improve the size of our NCRI signal we increased the
sample pressure toward 55 bar. The cell failed before reaching this
pressure.

\begin{figure}
\setlength{\unitlength}{1.0in}
\begin{picture}(2.5,2.5)
\put(0,0) { \makebox(2.5, 2.5)[t] {
\includegraphics[width=0.5\textwidth]{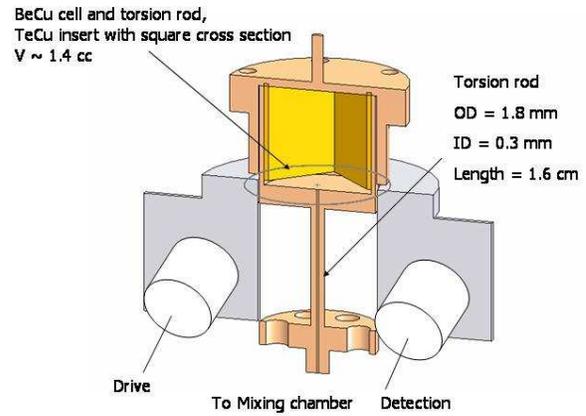}}
}
\end{picture}
\caption{\label{1oscillator}Torsional oscillator: The torsion cell's
motion is excited and detected capacitively. The AC voltage on the
detection electrodes serves as reference signal for a lock-in
amplifier to keep the oscillation in resonance. At 4 K, the
mechanical quality factor is 9 x 10$^5$, and the resonance frequency
is 185 Hz.}
\end{figure}

A diagram of our second cell is shown in Fig.\ \ref{1oscillator}.
The design is similar to that of our first cell with several
significant modifications. First, we have converted the interior of
the cell to a nearly cubic geometry by epoxying a Tellurium-Copper
insert into the cylindrical volume of the Beryllium-Copper
oscillator. The insert was machined to provide a cubic sample
region. We incorporated a capacitance strain gauge at the top of the
cell which allows us to determine the pressure after solidification.
In addition we have mounted a heater and thermometer directly on the
cell. The solid samples were grown with the blocked capillary
method, where the helium crystal was formed in the fill line first
while cooling. The heater on the cell gives us the option to melt
the sample while keeping the helium in the fill line and heat sinks
frozen. These changes have resulted in an increase of the moment of
inertia of the cell and reduction of the resonance frequency from
253 to 185 Hz.

We were motivated to adopt the noncircular sample geometry by the
recent suggestion of Dash and Wettlaufer \cite{4} that the reduction
in the moment of inertia observed by KC might be caused by slippage
of the solid, owing to grain boundary premelting at the surface of
contact between the solid helium and the walls of the cylindrical
container. The square cross section of our cubic sample cell
provides a geometric locking of the solid to the oscillating
container, so the mechanism of Dash and Wettlaufer \cite{4} may not
be invoked to explain the presence of a NCRI signal. Another
possibility is that the NCRI signal results from superflow along the
connected grain boundaries of a highly polycrystalline $^4$He solid.
Our observations cannot directly distinguish between this
possibility and a more homogeneous flow throughout the solid.

Much of the discussion of the supersolid state is premised on the
hypothesis that Bose-Einstein condensation (BEC) underlies the
supersolid phenomenon. For a BEC superfluid, the superfluid velocity
is determined by the gradient of the phase of the condensate
wavefunction; for incompressible superflow, the condensate phase
constitutes a velocity potential function satisfying Laplace's
equation. Fetter \cite{5} has considered the problem of superfluids
within rotating cylinders of various cross-sections. In the case of
a circular cross-section, the superfluid can remain at rest while
the wall rotates; in the case of a noncircular cross-section,
however, this is not possible because part of the fluid is displaced
by the walls as the cell rotates. In the case of a square
cross-section, Fetter finds a fraction of 0.156 times the solid body
moment of inertia of the fluid is entrained by the motion of the
walls. The tangential velocity of the superflow relative the walls
is nonuniform reaching a maximum value, v$_{max}$ = 1.35 L$\omega$/2
\cite{11}, at the center of the side walls, where $\omega$ is the
angular velocity and L is the edge length of the square cell. In
estimating the rim velocities we take this factor into account. The
corners of the square are stagnation points where the velocity of
the fluid relative to the wall is zero and the fluid moves with the
container.

\begin{figure}
\includegraphics[width=0.5\textwidth]{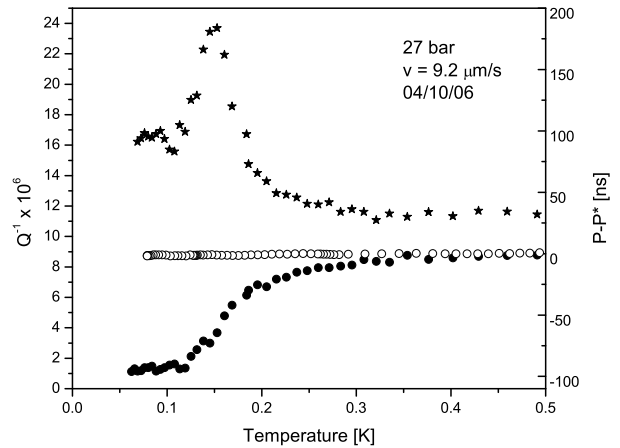}
\caption{\label{3pq}The resonant period (closed circles) and
dissipation data (stars) are shown as a function of temperature for
a sample formed at 27 bar. The period is shifted by 5.428053 ms, the
resonant period at 500 mK. The period of the empty cell (open
circles) is also adjusted to be zero at 500 mK. The maximum wall
velocity below 150 mK was 9.2 $\mu$m/s.}
\end{figure}

In Fig.\ \ref{3pq}, we show the period (closed circles) and
dissipation data, Q$^{-1}$(stars), plotted against cell temperature
for a typical measurement showing supersolid behavior. This sample
was formed by entering the solid phase at 27 bar. It then took 70
minutes to cool to 1 K. The data displayed were taken as we cooled
at a rate of 100 mK/hour to the lowest temperature. The maximum
velocity at the center of the walls below the transition temperature
is 9.2 $\mu$m/s. For comparison, we show the period data for the
empty cell (open circles) set to zero at 0.5 K. We expect the period
to be temperature independent for a system displaying classical
rotational inertia. The period data of our 27 bar sample follow CRI
down to a temperature near 0.3 K. As the temperature is reduced the
period gradually falls below the CRI value and changes most rapidly
below 0.2 K, where a marked peak appears in the dissipation. These
features are the NCRI signals as seen in the Penn State measurements
and signify the transition to the supersolid state. The sample was
cooled through the bcc into the hcp phase, which may explain the
large signal seen for this sample. These data bear a resemblance to
the torsional oscillator period and dissipation data seen for the
Kosterlitz-Thouless transition in $^{4}$He films \cite{6}.

The scale for the dissipation is determined by a ring-down
measurement of the quality factor. For the data shown in Fig.\
\ref{3pq}, the dissipation level, Q$^{-1}$, at temperatures above
the transition is on the order of 1.1 x 10$^{-5}$, to be compared to
an empty cell value of about 1.0 x 10$^{-6}$ at the same
temperature. For this sample, the solid contributes a substantial
excess dissipation in addition to that of the empty oscillator, even
at temperatures well above the supersolid transition. As will be
shown, this additional contribution to the dissipation of the
oscillator is largely absent for samples that do not show the
supersolid phenomenon.

\begin{figure}
\includegraphics[width=0.5\textwidth]{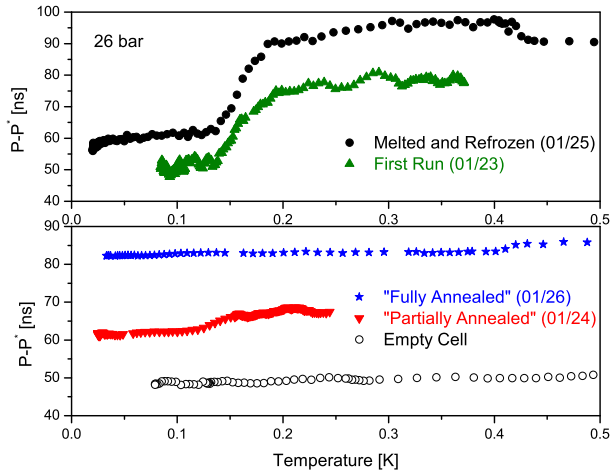}
\caption{\label{4period}The period data are shown as a function of
temperature for samples with different preparations. The data are
shifted by 5.417 ms. The empty cell period is additionally shifted
in order to be displayed in the same figure. The "partially
annealed" sample was held between 1.1 K and 1.2 K for 5 hours,
whereas the "fully annealed" sample was held between 1.4 and 1.5 K
for 13 hours before cooling down again. The period drop of these
samples is reduced or absent.}
\end{figure}

\begin{figure}
\includegraphics[width=0.5\textwidth]{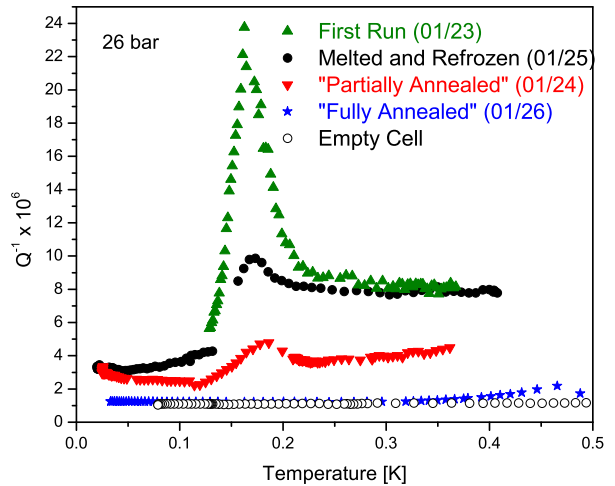}
\caption{\label{5dissipation}Dissipation data are shown as a
function of temperature for samples with four different sample
preparations. The samples showing a NCRI signal have dissipations on
the order of $8 \times 10^{-6}$ above the transition temperature and
display a dissipation peak, while the ``partially" annealed sample
shows a reduced dissipation above the NCRI transition temperature
and also a reduction the amplitude of the NCRI dissipation peak. The
``fully" annealed sample shows a lower dissipation of $1.2\times
10^{-6}$ at temperatures above the NCRI transition temperature and
no evidence of a peak in the dissipation at any temperature.}
\end{figure}

In Fig.\ \ref{4period} and \ref{5dissipation}, we show the period
and dissipation data for a sequence of runs which illustrate the
effect of annealing on the supersolid signal. The first run, on
01/23/06, shows the characteristic indication of the supersolid
transition in both the period and the dissipation signals. After
solidification, the cell was cooled at a rate of 150 mK/hour.
Between this run and the next, on 01/24/06, the cell was taken to a
temperature between 1.1 and 1.2 K, still below the melting
temperature, and held there for about 100 minutes and then allowed
to cool at a rate similar to that on 01/23. This ``partial"
annealing produced a marked reduction in the size of the NCRI period
signal, and there was also a reduction in the magnitude of the
dissipation peak.

Following this run, we heated the sample and the mixing chamber to a
temperature well above the melting temperature and then cooled
 in 30 minutes from the melting temperature of about 1.5 K  back
to below 1 K before beginning the run of 01/25. This rapid cool-down
produced a sample in which the full supersolid signal as seen in the
01/23 run was recovered.

In preparation for the next run on 01/26, we again annealed the
sample. The sample was held between 1.4 and 1.5 K for 13 hours
before being allowed to cool to near 0.5 K. Following this ``full"
annealing process, a low temperature sweep revealed no hint of a
supersolid state in either the period or in the dissipation data.

\begin{figure}
\includegraphics[width=0.5\textwidth]{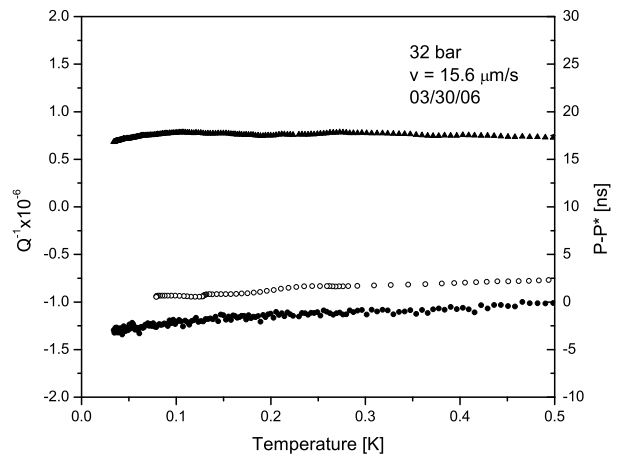}
\caption{\label{632bar}Resonance period (closed circles) and
dissipation (triangles) are shown as a function of temperature for a
sample formed at a pressure of 32 bar. The period is shifted by
5.428185 ms which is the resonance period at 300 mK. The empty cell
period (open circles) is shifted by 5.41704 ms, in order to be
displayed in the same panel. The duration of the cooldown (500 mK to
30 mK) was 7.5 hours. The maximum wall velocity of the oscillator at
80 mK is 15.6 $\mu$m/s.}
\end{figure}
We have also examined a few samples formed at pressures between 31.5
bar and 34.5 bar with the intention of avoiding any complications
associated with the bcc phase. The higher pressure samples all show
the CRI state. This may be due to annealing while cooling from the
higher temperatures required for entering the hcp solid phase at
pressures above the bcc phase. In Fig.\ \ref{632bar}, we show
Q$^{-1}$ (triangles) and P (closed circles) of such a sample in
which the solid phase was entered at a pressure of 32 bar. After
solidification it took 40 minutes to cool below 1 K. The dissipation
of 0.7 x 10$^{-6}$ is indistinguishable from the empty cell value
for this run. The period drops gradually by 3 ns without a distinct
drop. We melted and refroze the sample twice without observing the
NCRI state. We also repeated these measurements for maximum wall
velocities of 13.6 $\mu$m/s, 15.6 $\mu$m/s, 20.8 $\mu$m/s, and 36.6
$\mu$m/s with the same result.

\begin{table*}
\caption{\label{tab:table1}The values for the maximum wall velocity,
dissipation peak temperature, dissipation at 300 mK, and the
observed period shift at the NCRI transition temperature, are listed
for the data sets displayed in the Figures 2 - 5. The velocities are
calculated at 80 mK,  and at 300 mK. The critical velocities
according to \cite{2} are 38 $\mu$m/s and 5 $\mu$m/s for 26 bar and
32 bar respectively. The empty cell dissipation at 300 mK is 1.1 x
10$^6$.}
\begin{ruledtabular}
\begin{tabular}{lccccccc}
&p&Date& v$_{80 mK}$ & v$_{300 mK}$ & T(Q$^{-1}_{max}$)&Q$^{-1}$ x 10$^6$(300 mK)&$\Delta$P\\
&[bar]&&[$\mu$m/s]&[$\mu$m/s] & [mK]&&[ns]\\
First Run &26& 01/23& 25.4 & 36.6 & 163 &8.2& 27\\
Partial Anneal &26 & 01/24& 39.6 & 36.7 &186&3.9& 6\\
Melted/Refrozen &26& 01/25& 16.3&  12.9& 170&7.9& 35\\
Full Anneal &26 &01/26& 45.7 &  44.4&N/A &1.2&0\\
&27&03/30 & 9.2 & 11.3 &148&11.0& 94\\
&32 &04/10 & 15.6 & 17.2& N/A & 0.7&0\\
\end{tabular}
\end{ruledtabular}
\end{table*}

In table \ref{tab:table1}, we show a summary of velocity and period
shift data for the runs shown in Figures \ref{3pq}, \ref{4period},
\ref{5dissipation}, \ref{632bar}. According to KC \cite{2}, the
critical velocity at 26 bar in an annulus is 38 $\mu$m/s and the
period drop is reduced by about 35 \% at 65 $\mu$m/s. With maximal
velocities of 39.6 $\mu$m/s, and 45.7 $\mu$m/s, we would expect a
period drop of at least 20 ns, a reduction by 35 \%. At 30 bar, KC
\cite{10} find a signal reduction of about 30 \% for a velocity of
27 $\mu$m/s, which cannot explain the absence of NCRI in our
experiments; v$_{80 mK}$ is 13.6 $\mu$m/s.

An important aspect of the data is the correlation between CRI and
low dissipation levels at 300 mK. We believe that the high Q found
for our non-supersolid samples indicates a state with fewer defects
than in the supersolid samples. Together with the metastability of
the NCRI, this observation suggests that crystal imperfections, such
as dislocations, vacancies, and grain boundaries, whose presence is
indicated by excess dissipation, are essential for the existence of
the supersolid state.

These findings are in accord with the recent
path-integral-monte-carlo calculations of B.K. Clark and D.M.
Ceperley \cite{7}, who find that an ideal hcp $^{4}$He crystal will
not support BEC or off-diagonal-long-range-order (ODLRO) and
consequently is not expected to display supersolid behavior.
Further, M. Boninsegni, N. Prokof'ev, and B. Svistunov \cite{8} find
that a degree of disorder is required for the $^{4}$He solid to
become a supersolid.

In this letter, we report two findings. First, we have observed the
supersolid signal for solid $^{4}$He samples contained in both
cylindrical and cubic geometries. The fractional period shift
observed is of the same order of magnitude as that reported by Kim
and Chan. This contrasts with the report of Keiya Shirahama (Keio
University, Japan) \cite{9}, who has observed a supersolid signal an
order of magnitude smaller than that seen in the Penn State and
Cornell experiments.

Second, we find that the supersolid signal is not a universal
property of solid $^{4}$He, but can be reduced and even eliminated
in annealed samples. These annealed non-NCRI samples display lower
dissipation levels than the NCRI samples, indicating, we believe, a
lower degree of crystal imperfection. The absence of supersolid
behavior in the annealed samples is supported by the recent
theoretical findings that an ideal hcp $^{4}$He crystal does not
support ODLRO or BEC \cite{7},\cite{8}. We note that the Penn State
and Keio University groups \cite{priv} have both undertaken
annealing experiments. Changes have been seen in the signal size;
however, these experiments have not been successful in eliminating
the NCRI signal.

\begin{acknowledgements}
We would like to acknowledge the support of our colleagues at
Cornell and elsewhere.  In particular, we thank Moses Chan, Erich
Mueller, J. Seamus Davis, Jeevak Parpia and David Lee for their
encouragement and assistance. We are indebted to Kelken Chang for
providing the detailed solution for the potential flow in a square
geometry. One of us, JDR, would like to thank the Kavli Institute
for Theoretical Physics and the Aspen Center for Physics for their
hospitality. The work reported here has been supported by the
National Science Foundation under Grant DMR-0203244 and through the
Cornell Center for Materials Research under Grant DMR-0520404.
\end{acknowledgements}


\begin{thebibliography}{11}
\expandafter\ifx\csname
natexlab\endcsname\relax\def\natexlab#1{#1}\fi
\expandafter\ifx\csname bibnamefont\endcsname\relax
  \def\bibnamefont#1{#1}\fi
\expandafter\ifx\csname bibfnamefont\endcsname\relax
  \def\bibfnamefont#1{#1}\fi
\expandafter\ifx\csname citenamefont\endcsname\relax
  \def\citenamefont#1{#1}\fi
\expandafter\ifx\csname url\endcsname\relax
  \def\url#1{\texttt{#1}}\fi
\expandafter\ifx\csname urlprefix\endcsname\relax\def\urlprefix{URL
}\fi \providecommand{\bibinfo}[2]{#2}
\providecommand{\eprint}[2][]{\url{#2}}

\bibitem[{\citenamefont{Kim and Chan}(2004{\natexlab{a}})}]{1}
\bibinfo{author}{\bibfnamefont{E.}~\bibnamefont{Kim}} \bibnamefont{and}
  \bibinfo{author}{\bibfnamefont{M.~H.~W.} \bibnamefont{Chan}},
  \bibinfo{journal}{Nature} \textbf{\bibinfo{volume}{427}},
  \bibinfo{pages}{225} (\bibinfo{year}{2004}{\natexlab{a}}).

\bibitem[{\citenamefont{Kim and Chan}(2004{\natexlab{b}})}]{2}
\bibinfo{author}{\bibfnamefont{E.}~\bibnamefont{Kim}} \bibnamefont{and}
  \bibinfo{author}{\bibfnamefont{M.~H.~W.} \bibnamefont{Chan}},
  \bibinfo{journal}{Science} \textbf{\bibinfo{volume}{305}},
  \bibinfo{pages}{1941} (\bibinfo{year}{2004}{\natexlab{b}}).

\bibitem[{\citenamefont{kelken}(2006{\natexlab{b}})}]{11}
\bibinfo{author}{\bibfnamefont{K.}~\bibnamefont{Chang}}
  \bibinfo{journal}{private communication}  (\bibinfo{year}{2006}).

\bibitem[{\citenamefont{Wong}(1988)}]{3}
\bibinfo{author}{\bibfnamefont{G.~K.} \bibnamefont{Wong}},
  \bibinfo{journal}{Torsional Oscillators in Experimental Techniques in Condensed Matter Physics at Low
  Temperatures, Eds. R. C. Richardson and E. N. Smith, Addison-Wesley,} \bibinfo{pages}{187} (\bibinfo{year}{1988}).

\bibitem[{\citenamefont{Dash and Wettlaufer}(2005)}]{4}
\bibinfo{author}{\bibfnamefont{J.~G.} \bibnamefont{Dash}} \bibnamefont{and}
  \bibinfo{author}{\bibfnamefont{J.~S.} \bibnamefont{Wettlaufer}},
  \bibinfo{journal}{Phys.\ Rev.\ Lett.} \textbf{\bibinfo{volume}{94}},
  \bibinfo{pages}{235301} (\bibinfo{year}{2005}).

\bibitem[{\citenamefont{Fetter}(1974)}]{5}
\bibinfo{author}{\bibfnamefont{A.~L.} \bibnamefont{Fetter}},
  \bibinfo{journal}{J. Low Temp. Phys.} \textbf{\bibinfo{volume}{16}}, \bibinfo{pages}{533}
  (\bibinfo{year}{1974}).

\bibitem[{\citenamefont{Bishop and Reppy}(1978)}]{6}
\bibinfo{author}{\bibfnamefont{D.}~\bibnamefont{Bishop}} \bibnamefont{and}
  \bibinfo{author}{\bibfnamefont{J.}~\bibnamefont{Reppy}},
  \bibinfo{journal}{Phys.\ Rev.\ Lett.} \textbf{\bibinfo{volume}{40}},
  \bibinfo{pages}{1727} (\bibinfo{year}{1978}).

\bibitem[{\citenamefont{Clark and Ceperley}(2004)}]{7}
\bibinfo{author}{\bibfnamefont{B.~K.} \bibnamefont{Clark}} \bibnamefont{and}
  \bibinfo{author}{\bibfnamefont{D.~M.} \bibnamefont{Ceperley}},
  \bibinfo{journal}{Phys.\ Rev.\ Lett.} \textbf{\bibinfo{volume}{93}},
  \bibinfo{pages}{155303} (\bibinfo{year}{2004}).

\bibitem[{\citenamefont{Boninsegni et~al.}(2006)\citenamefont{Boninsegni,
  Prokof'ev, and Svistunov}}]{8}
\bibinfo{author}{\bibfnamefont{M.}~\bibnamefont{Boninsegni}},
  \bibinfo{author}{\bibfnamefont{N.}~\bibnamefont{Prokof'ev}},
  \bibnamefont{and}
  \bibinfo{author}{\bibfnamefont{B.}~\bibnamefont{Svistunov}},
  \bibinfo{journal}{Phys.\ Rev.\ Lett.} \textbf{\bibinfo{volume}{96}},
  \bibinfo{pages}{105301} (\bibinfo{year}{2006}).

\bibitem[{\citenamefont{Shirahama}(2006)}]{9}
\bibinfo{author}{\bibfnamefont{K.}~\bibnamefont{Shirahama}},
  \bibinfo{journal}{BAPS, 2006 March Meeting,} \bibinfo{pages}{302}
  (\bibinfo{year}{2006}).

\bibitem[{\citenamefont{Chan and Shirahama}(2006)}]{priv}
\bibinfo{author}{\bibfnamefont{M.~H.~W.} \bibnamefont{Chan}} \bibnamefont{and}
  \bibinfo{author}{\bibfnamefont{K.}~\bibnamefont{Shirahama}},
  \bibinfo{journal}{private communication}  (\bibinfo{year}{2006}).

\bibitem[{\citenamefont{Kim and Chan}(2006{\natexlab{b}})}]{10}
\bibinfo{author}{\bibfnamefont{E.}~\bibnamefont{Kim}} \bibnamefont{and}
  \bibinfo{author}{\bibfnamefont{M.~H.~W.} \bibnamefont{Chan}},
  \bibinfo{journal}{cond-mat/0605680} (\bibinfo{year}{2006}{\natexlab{b}}).
\end{thebibliography}
\end{document}